\newif\ifeprint\eprinttrue
\ifeprint
\documentclass
[rmp,twocolumn,amsfonts,amsmath,amssymb,showkeys,letterpaper,raggedbottom]{revtex4}
\else
\documentclass
[rmp,preprint,amsfonts,amsmath,amssymb,showkeys,letterpaper,raggedbottom]{revtex4}
\fi
\date{\today}

\def\TITLE{Modeling molecules with constraints}
\def\KEYWORDS{molecular simulation, constrained moves, energy evaluation}

\usepackage{times}
\usepackage[bookmarks=false]{hyperref}
\hypersetup{pdftitle={\TITLE},
pdfauthor={C. F. F. Karney; J. E. Ferrara},
pdfkeywords={\KEYWORDS}}

\ifeprint
\else
\fi
\bibpunct{[}{]}{;}{n}{}{,}

\makeatletter\def\raggedcolumn@skip{\vskip\z@\@plus.0001fil\relax}\makeatother

\def\hrefx#1#2#3{\href{#1}{#2}\penalty0\href{#1}{#3}}

\DeclareMathAlphabet{\mathbsf}{OT1}{cmss}{bx}{n}
\def\v#1{\mathbf{#1}}
\def\abs#1{\left|#1\right|}
\def\stack#1#2{\genfrac{}{}{0pt}{2}{#1}{#2}}

\begin{document}

\ifeprint
\noindent\mbox{\begin{minipage}[b]{\textwidth}
\begin{flushright}\tt\footnotesize
Link: \href{http://charles.karney.info/biblio/constraint.html}
           {http://charles.karney.info/biblio/constraint.html}\par\vspace{0.5ex}
E-print: \href{http://arxiv.org/abs/physics/0508116}
                             {arXiv:physics/0508116}\par
\vspace{2ex}
\end{flushright}
\end{minipage}
\hspace{-\textwidth}}
\fi

\title{\TITLE}
\ifeprint
\author{\href{http://charles.karney.info}{Charles F. F. Karney}}
\else
\author{Charles F. F. Karney}
\fi
\email{ckarney@sarnoff.com}
\author{Jason E. Ferrara}
\affiliation{\href{http://www.sarnoff.com}{Sarnoff Corporation},
  Princeton, NJ 08543-5300}

\begin{abstract}
Techniques for simulating molecules whose conformations satisfy
constraints are presented.  A method for selecting appropriate moves in
Monte Carlo simulations is given. The resulting moves not only obey the
constraints but also maintain detailed balance so that correct
equilibrium averages are computed.  In addition, techniques for
optimizing the evaluation of implicit solvent terms are given.

\end{abstract}

\maketitle

\section{Introduction}

When attempting to compute thermodynamic quantities with a molecular
simulation, we are frequently confronted with the problem of sampling in
a high-dimensional configuration space.  The dimensionality of this
space is given by the number of degrees of freedom for the molecular
system.  Techniques which lower the number of degrees of freedom will
increase the efficiency of the thermodynamic sampling---provided, of
course, that these techniques are physically justified.  Thus, an
implicit solvent model may be used to eliminate the degrees of freedom
associated with the solvent molecules; the standard chemical force
fields replace the electron charges with atom-centered partial charges
thereby removing the electrons' degrees of freedom.  Further reductions
in dimensionality are possible by imposing constraints on the relative
positions of the atoms in a molecule.  Thus we might specify that the
bond lengths and bond angles in a molecule are fixed and only the
torsion angles are allowed to vary.  It is such a scenario that we
examine in this paper.  We address two aspects of this problem: how to
move a molecule subject to constraints in order to allow equilibrium
averages to be computed using the canonical-ensemble Monte Carlo method
\cite{metropolis53} and how to evaluate the energy efficiently.

The imposition of constraints in molecular modeling has been extensively
studied \cite[\S3.3.2,~\S15.1]{frenkel02}.  Let us start by elucidating
the difference in the treatment of hard constraints in molecular
dynamics and Monte Carlo simulations.  We treat hard constraints by
taking the limit where the ``spring constant'' for the hard degrees of
freedom is infinite.  Molecular dynamics simulations then consider the
evolution of the resulting system over a finite time.  On the other
hand, if we wish to determine the equilibrium properties of a system
using the Monte Carlo method, then we need to consider averaging over
sufficiently long times to allow equipartition of energy among all the
degrees of freedom of a system.  This is, of course, an example of
nonuniform limits.  We are interested in taking both $\tau_\mathrm{sim}
\rightarrow \infty$ and $\tau_\mathrm{equ} \rightarrow \infty$, where
$\tau_\mathrm{sim}$ is the representative simulation time and
$\tau_\mathrm{equ}$ is the equipartition time (which is proportional to
the ``stiffness'' of the constraints).  In constrained molecular
dynamics, we take the limit $\tau_\mathrm{equ} \rightarrow \infty$
first, which prevents equipartition from occurring; whereas in
equilibrium statistical mechanics, we take $\tau_\mathrm{sim}
\rightarrow \infty$ first and this allows energy equipartition.  If we
are attempting to compute an equilibrium quantity, such as the free
energy of binding, it is essential to allow energy equipartition.
Understanding this distinction explains the apparently contradictory
results for constrained and unconstrained averages for a flexible trimer
\cite[\S15.1]{frenkel02}.

One way of understanding the constrained equilibrium system is to
consider how the equilibrium distribution varies as the constraint is
imposed.  In the limit, the distribution collapses to a
lower-dimensional sub-manifold of configuration space.  However, this
sub-manifold has a ``thickness'' that depends on the details of the
constraint term and, consequently, Monte Carlo moves for the constrained
system need to reflect this thickness in order to sample the
distribution correctly.  As a consequence, we will need to specify the
functional form of the constraint energy and the constraint is no longer
a purely geometrical object.  At first glance, this would appear to
complicate further the already complex algebra of constrained motions
\cite{fixman74}.  However, we will propose an algorithm for making moves
which is simple to implement and which automatically ensures that the
correct equilibrium averages are computed.

The second half of the paper considers a mundane---but nevertheless
important---problem, namely how to evaluate the energy of a molecule
made up of rigid subcomponents.  We propose a consistent framework for
avoiding the computation of constant terms and for imposing energy
cutoffs.  We extend this to the computation of the generalized Born
solvation term and we describe a simple method for computing the solvent
accessible surface area which has a bounded error.

\section{Generalized Monte Carlo moves}

We begin by assembling some techniques for combining Monte Carlo moves.
We define an ``$E$ move'' as an ergodic move which preserves
$\exp(-\beta E)$ as the invariant distribution, where $\beta = 1/(kT)$
and $k$ is the Boltzmann constant.  (Here, ``ergodic'' implies that the
move allows all relevant portions to configuration space to be
explored.)  Note that a zero move has a uniform invariant distribution.
A typical zero move samples a new configuration from a distribution
which satisfies the symmetry requirement $p(\Gamma';\Gamma) =
p(\Gamma;\Gamma')$, where $p(\Gamma';\Gamma)$ is the probability density
of picking a new configuration of $\Gamma'$ given a starting
configuration of $\Gamma$.  Clearly a sequence of $n$ $E$ moves is
itself an $E$ move.  From the central limit theorem, a sequence of $n$
zero moves is equivalent, in the limit of large $n$, to selecting the
new configuration from a multi-dimensional Gaussian.

Instead of carrying out the $n$ moves with a given energy $E(\Gamma)$,
we can consider the case where the energy is given by
$E_\lambda(\Gamma)$ which depends continuously on the parameter
$\lambda$.  A sequence of $n$ $E_\lambda$ moves where $\lambda$ is
varied {\em adiabatically} in such a way that its initial and final
values are $\lambda_0$ is an $E_{\lambda_0}$ move.  This follows because
adiabatically varied systems are always in equilibrium with the
instantaneous value of $\lambda$ \cite[\S11]{landau69}.  Each
$E_\lambda$ move is carried out at a {\em fixed} $\lambda$ and $\lambda$
is varied between the moves.  In order to satisfy the adiabatic
condition, we will need to take $n$ large.

A move from $\Gamma$ to $\Gamma'$ may be subjected to ``Boltzmann
acceptance with energy $E$''.  This involves accepting the move
($\Gamma'$ is the new state) with probability $M(x)$ and otherwise
rejecting the move ($\Gamma$ is the new state).  Here $M(x)$ is a
function satisfying $0 < M(x) \le 1$ and $M(x)/M(-x) = \exp(-x)$ with $x
= \beta(E(\Gamma') - E(\Gamma))$.  Usually we take $M(x) =
\min(1,\exp(-x))$; however other choices, e.g., the Fermi function,
$M(x) = 1/(1+\exp(x))$, are possible \cite{bennett76}.

Consider an $E_1$ move from $\Gamma$ to $\Gamma'$ followed by an
Boltzmann acceptance using $E_2$.  This compound move is an $(E_1 +
E_2)$ move.  The proof follows as a special case of the ``multiple
time-step'' (MTS) method \cite[\S II]{hetenyi02} or, alternatively, as a
special case of early rejection \cite[\S14.3.2]{frenkel02}.  If the
$E_1$ move was already a ``rejected'' move, i.e., $\Gamma'=\Gamma$, then
the Boltzmann test involving $E_2$ automatically ``succeeds'' ($M(0) =
1$).  Thus $E_2$ does not need to be evaluated in this case.

These results allow us to generalize the MTS method by splitting the
energy into $m$ terms (instead of just two),
\[
E(\Gamma) = \sum_{l=1}^m E_l(\Gamma).
\]
The method is defined recursively as follows: a level-$0$ move is
defined to be a zero move; a level-$l$ move, with $l>0$, is defined to
be $n_{l-1}$ level-$(l-1)$ moves the result of which is subjected to
Boltzmann acceptance using $E_l$.  By induction, we see that a level-$l$
move is an $\mathcal E_l$ move, where
\[
\mathcal E_l(\Gamma) = \sum_{l'=1}^l E_{l'}(\Gamma).
\]
It follows that a level-$m$ move is an $\mathcal E_m$ move, i.e., an $E$
move.  Typically we sample the zero moves from a Gaussian and we take
$n_0 = 1$.  Standard Monte Carlo \cite{metropolis53} is given by $m=1$
and $n_1=1$.  Standard MTS \cite{hetenyi02} is recovered with $m=2$.
The early rejection method \cite[\S14.3.2]{frenkel02} is recovered with
$n_l=1$ (for all $l$).  Note that a level-$m$ move entails
$\prod_{l'=l}^{m-1} n_{l'}$ level-$l$ moves.  At any stage in the
recursion, we have the freedom to vary some of the components of
$E(\Gamma)$ adiabatically.

In the following sections, we apply these techniques to constrained
molecules.  In simple cases, we can apply the MTS method
semi-analytically to derive a correct constrained move.  In more
complicated cases, we apply the adiabatic technique to lift and to
reapply the constraint.

\section{Stiff molecules}

A constrained molecule is a mathematical idealization of a real system
in which some degrees of freedom are stiff, i.e., the associated
energies are large.  Thus we can split the energy into ``hard''
($\mathrm h$) and ``soft'' ($\mathrm s$) components,
\[
E(\Gamma) = E_\mathrm h(\Gamma) + E_\mathrm s(\Gamma),
\]
where $\Gamma$ is the configuration of the system.  For example, let us
assume that an all-atom force field, such as Amber \cite{cornell95},
provides an accurate description of the system.  (We recognize, of
course, that present-day force fields are only approximate.  However,
our purpose here is to make the connection between an all-atom
representation and a simpler rigid representation and, in this context,
the details of the all-atom model are of secondary importance.)  Then
$E_\mathrm h$ might represent the bond stretching and bond bending
terms, while $E_\mathrm s$ is given by the other terms (bond torsion and
the non-bonded energies).

The constrained limit is now given by $E_\mathrm h \rightarrow \infty$.
Before we consider this limit, it is useful to examine how the stiff
system may be treated.  Conventional Monte Carlo is inefficient because,
in order to have an reasonably large acceptance rate, the step-size
needs to be set to a small value (determined by $E_\mathrm h$) so that
diffusion in the soft directions is very slow.  However, we can apply
MTS Monte Carlo in this case with $E_1 = E_\mathrm h$ and $E_2 =
E_\mathrm s$.

Let us apply this method to a system of ``rigid'' molecules, e.g., water
molecules, taking $E_\mathrm h$ to include the intra-molecular energies
(responsible for maintaining the rigidity) and $E_\mathrm s$ to include
the inter-molecular energies.  Suppose the level-$0$ moves consist of
symmetrically displacing the atoms in each molecule.  The result of the
$n_1$ level-1 Monte Carlo steps will clearly be a symmetric,
independent, and nearly rigid displacement (translation and orientation)
of each molecule.  This configuration is then subjected to Boltzmann
acceptance with the inter-molecular energies.  In this case, we can
easily pass to the constrained limit (with exact rigidity), merely by
ensuring that the trial (level-1) moves of the molecules are rigid.  In
this case, we have just rederived the ``standard'' move for a system of
rigid molecules.

In order to illustrate the application to flexible molecules, we shall
treat the molecules as being made up of several rigid subunits or
``fragments'' connected by flexible bonds.  However we are interested in
the limit where the inter-fragment bonds constrain the relative motions
of fragments in certain ways, either by fixing the bond lengths
(allowing the bond angles and bond dihedrals to vary) or by fixing the
bond lengths and bond angles (allowing the bond dihedrals to vary).
Such a model is adequate to describe a wide range of interesting organic
molecules including proteins and drug-like ligands.  We assume that the
rigidity of the fragments is imposed only by intra-fragment energy.  If
other terms (e.g., an improper torsion term involving atoms from two
fragments) contribute to the rigidity of a fragment, then we shall treat
such terms as additional inter-fragment energies.

We apply the generalized MTS method to this system with $m=3$, the
intra-fragment energy given by $E_1$, the inter-fragment bond
constraints given by $E_2$, and with $E_3$ accounting for all the other
energies.  The argument given above allows us to pass to the limit of
strictly rigid fragments.  The method is then equivalent to a standard
MTS method where the ``elementary'' moves consist of rigid displacements
of each fragment which are Boltzmann accepted with energy $E_\mathrm h =
E_2$.  A sequence of $n = n_2$ such moves are made with the result
Boltzmann accepted with energy $E_\mathrm s = E_3$.  A possible
prescription \cite[\S VII]{karney05b} for the rigid displacements of the
fragments is to translate the fragment by a vector sampled from an
isotropic 3-dimensional Gaussian and to rotate the fragment by $\abs{\v
s}$ about an axis $\hat{\v s}$ where $\v s$ is a ``rotation vector''
also sampled from an isotropic 3-dimensional Gaussian.  The variances
for the two Gaussians should be adjusted so that the translational and
rotational components result in comparable displacements of the atoms of
the fragment.

Provided that the inter-fragment constraint terms $E_\mathrm h$ are
sufficiently stiff, it is not important to include a detailed model of
these terms; because the motion will take place near the bottom of the
constraint potential well, a harmonic (i.e., quadratic) approximation to
the constraint potential will suffice.  On the other hand, if the
stiffness of the constraint energy depends on any of the soft degrees of
freedom, it is important that this effect be included.

It is frequently the case that $E_\mathrm h$ may be computed much more
rapidly than $E_\mathrm s$.  For example, when imposing bond constraints
on a molecule, $E_\mathrm h$ requires $O(N)$ computations, where $N$ is
the number of atoms, while $E_\mathrm s$ requires $O(N^2)$ computations
for the electrostatic and implicit solvation energies.  Thus we might be
able to take $n$ reasonably large and still have the computational cost
dominated by the evaluation of $E_\mathrm s(\Gamma)$.

In order to realize the full benefits of imposing constraints we need to
pass to the constrained limit ($E_\mathrm h \rightarrow \infty$).  In
this limit, the motion collapses onto a lower-dimensional sub-manifold
in configuration space.  Unfortunately, in contrast to the case of rigid
molecules, we cannot appeal to symmetry to enable us to take this limit
analytically.  Instead, we use the adiabatic technique.

\section{Adiabatically varying the stiffness}

Let us rewrite the energy of the system, multiplying the $E_\mathrm
h(\Gamma)$ by $T/T^*$, where $T$ is the temperature of the system, and
$T^*$ is a ``constraint'' temperature.  The Boltzmann factor
$\exp(-\beta E)$, will then have the form
\[
\exp(-\beta E) = \exp(-\beta E_\mathrm s - \beta^* E_\mathrm h)
\]
where $\beta^* = 1/(kT^*)$.

In our application, where we are interested in the constrained limit
$T^*\rightarrow 0$, a direct application of the MTS method leaves us
with two bad choices.  If we take $T^*$ to be sufficiently small that we
can consider the constraints to be satisfied, we will have to chose the
step size for the $E_\mathrm h$ moves to be so small that the change in
configuration after $n$ $E_\mathrm h$ moves will be small.  On the other
hand, letting $T^*$ be sufficiently large to allow moves will result in
configurations where the constraints are poorly satisfied.

We overcome this difficulty by regarding $T^*$ as a parameter (taking
the place of $\lambda$) and by adiabatically varying $T^*$ from zero
(where the constraints are satisfied but MTS is ineffective at making
moves) to a finite value (where the constraints are relaxed and MTS
becomes effective) and back to zero again (to reimpose the constraints).
During the course of changing $T^*$, we make $n$ $E_\mathrm h$ moves
(each with the instantaneous value of $T^*$).  The effect of these $n$
moves will be an $E_\mathrm h$ move with $T^* = 0$, i.e., a move which
satisfies the $E_\mathrm h$ constraint.

It remains to give a recipe for varying $T^*$.  As we vary $T^*$, we
would naturally adjust the step size for the moves in such a way that
the number of steps needed to equilibrate the system is a constant,
suggesting that we vary $T^*$ exponentially.  We therefore pick
\[
T^*_i = \left\{
\begin{array}{l@{\hspace{1em}}l}
T^*_A \exp(\alpha (i - 1)), & \mbox{for $0 < i \le m$},\\
T^*_A \exp(\alpha (n - i)), & \mbox{for $m < i \le n$},
\end{array}\right.
\]
where we have taken $n = 2m + 1$ and where $T^*_i$ is the constraint
temperature used for the $i$th $E_\mathrm h$ move, $T^*_{0} = T^*_A$ is
some temperature sufficiently small that we can consider the constraints
to be exactly satisfied, and $\alpha$ is the rate of increase of the
temperature which should be sufficiently small that the adiabatic
condition is satisfied.  Even though $T^*_A$ and $\alpha$ are small, we
can pick $n$ sufficiently large that $T^*_{m+1} = T^*_B = T^*_A
\exp(\alpha m)$ is finite.

In addition, we choose the step size for the $i$th $E_\mathrm h$ move to
be $d_i = k \sqrt{T^*_i}$ where $k$ is a constant.  In traditional Monte
Carlo, we normally pick $k$ to maximize the diffusion rate which at the
$i$th step is roughly
\[
    D_i = \frac{\langle (\Gamma_i - \Gamma_{i-1})^2\rangle}2
\sim \frac12 A d_i^2,
\]
where $A$ is the mean acceptance rate and $\langle\ldots\rangle$ denotes
an ensemble average.  Maximizing the diffusion rate usually results in a
rather small acceptance rate $A \sim 0.1$ because rare large steps can
lead to faster diffusion than frequent small steps.  However, in our
application, where we want the system to remain in equilibrium as we
vary the temperature, rare large steps are {\em bad}.  So we pick
$k$ to maximize $A D_i$ and this will usually result in $A \sim 0.5$.
Note that for a given $k$, we have
\[
    D_i \sim C T^*_i,
\]
where $C$ is constant provided that the step size is not too large.  The
overall diffusion can be estimated by summing over the $n$ steps,
\[
    D =  \frac{\langle (\Gamma_n - \Gamma_0)^2\rangle}2 =
\sum_{i=1}^n D_i \sim 2 C T^*_B/\alpha,
\]
where we have assumed that successive steps are uncorrelated and we have
taken $\alpha\ll 1$ and $T^*_B \gg T^*_A$.  We should select parameters,
$\alpha$ and $T^*_B$, in order to adjust $D$ so that the $E_\mathrm s$
acceptance rate is $O(1)$.

This method includes internal diagnostics to verify that $\alpha$ is
small enough.  We define
$\overline{\vphantom{A}\ldots}\mathord{\uparrow}$
(resp.~$\overline{\vphantom{A}\ldots}\mathord{\downarrow}$) as the
average of a quantity over the steps where $T^*_i$ is increasing, i.e.,
$i \le m+1$ (resp.~decreasing, i.e., $i> m+1$).  We monitor
$\overline{E_\mathrm h(\Gamma_i)/T^*_i}\mathord{\uparrow}$ and
$\overline{E_\mathrm h(\Gamma_i)/T^*_i}\mathord{\downarrow}$ and demand
that both should be close to the equilibrium value of $N/2$ (where $N$
is the number of hard degrees of freedom).  If $\alpha$ is too large,
then we would find
\begin{eqnarray*}
\overline{ E_\mathrm h(\Gamma_i)/T^*_i }\mathord{\uparrow} &\ll& N/2, \\
\overline{ E_\mathrm h(\Gamma_i)/T^*_i }\mathord{\downarrow} &\gg& N/2.
\end{eqnarray*}
In particular, if the final $E_\mathrm h(\Gamma_n)$ is many times
$T^*_A$, then the configuration is ``hung up'' and does not obey the
constraints.  If this happens frequently, the simulation needs to be
rerun with a smaller setting for $\alpha$; if, on the other hand, it
happens only rarely, we would merely reject the step.  We can also
monitor the mean acceptance rates $ \overline A \mathord{\uparrow}$ and
$\overline A \mathord{\downarrow}$.  These should be about the same;
however, if $\alpha$ is too large, we will find $ \overline A
\mathord{\uparrow} \gg \overline A \mathord{\downarrow}$.

A useful guideline for picking $T^*_A$ is that once the $n$ $E_\mathrm
h$ moves are completed and the system is presumably equilibrated to
$T^*_A$, we should be able to enforce the constraints by setting $T^* =
0$ (using any convenient energy minimization technique) with a
negligible change in the configuration, e.g., with a negligible change
in $E_\mathrm s(\Gamma)$.

\section{Pairwise terms in energy}

Having made an adiabatic move using $E_\mathrm h$, the final step is to
accept the move depending on the change in $E_\mathrm s$.  We wish to
compute this energy as efficiently as possible by using the rigidity of
the fragments.  Force fields such as Amber \cite{cornell95} include two
types of energies: interactions between atoms (the electrostatic and
Lennard-Jones terms) and bond energies (stretch, bend, and torsion).
Since the number of terms in non-bonded energies typically scales as
$O(N^2)$ where $N$ is the total number of atoms in the system, while the
number of bond terms scales as $O(N)$, we concentrate on optimizing the
evaluation of the non-bonded terms.  In our case where the molecules
consist of rigid fragments connected by flexible bonds we need only
include the bond terms contributed by the much smaller number of
inter-fragment bonds.  Furthermore, we need only include the energy
contributed by the ``free'' components of such bonds.  Thus, if the
lengths and angles of such bonds are constrained, then we need only
include the torsion energy in $E_\mathrm s(\Gamma)$.

We start by assuming that the non-bonded energy terms can be expressed
as a sum over atom pairs.  This applies to the electrostatic and
Lennard-Jones terms in Amber \cite{cornell95}.  However, implicit
solvent models have a more complex structure and we consider these in
the next section.

Suppose our molecular system consists of $N$ atoms.  These atoms are
grouped into $M$ molecules and we denote $M_l$ as the set of atoms
making up the $l$th molecule.  Similarly, the atoms are divided into $F$
rigid fragments and we denote $F_a$ as the set of atoms making up the
$a$th fragment.  A typical pairwise energy term can then be written as
\[
E_g(\Gamma) = \sum_{0< i<j \le N} C_{g,ij} f_g(r_{ij}),
\]
where $g$ denotes the type of energy term (electrostatic or
Lennard-Jones), $i$ and $j$ are atom indices, $r_{ij}$ is the distance
between atoms $i$ and $j$, $f_g$ is some function of distance, and
$C_{g,ij}$ is a coefficient which depends on the atoms but not on their
positions.  Thus for electrostatic interactions, $C_{g,ij}$ depends on
the partial charges on the two atoms (assumed to be constant in Amber)
and on the bonding relation between the atoms.  Physical energy
functions satisfy $\lim_{r\rightarrow\infty}f_g(r) = 0$.  When the
fragments are separated sufficiently, we have
\[
E_g \rightarrow E_{g0} =
 \sum_{0< a \le F} \sum_{\stack{i<j}{i,j \in F_a}} C_{g,ij} f_g(r_{ij}),
\]
which is independent of $\Gamma$.  It is convenient to choose $E_{g0}$
as the ``origin'' for the $E_g$, i.e., we compute only
\[
E_{g1} = E_g - E_{g0} = \sum_{0< a<b \le F}
\sum_{\stack{i \in F_a}{j \in F_b}}
C_{g,ij} f_g(r_{ij}).
\]
We note that only energy differences enter into the computation of
observable quantities, and so we are free to select the arbitrary origin
for energies.  

Let us consider the application of a small molecule ($N_l$ atoms)
interacting with a protein ($N_p \gg N_l$ atoms) where only some of the
protein side chains near the binding site are allowed to move.  By
avoiding computing the interaction energy between atoms in the immobile
portion of the protein, the above prescription reduces the computational
cost from $O(N_p^2)$ to $O(N_l N_p)$.

This cost may still be too large and we can substantially reduce the
cost by implementing energy cutoffs for the interactions.  This is
easily accomplished by multiplying $f_g(r_{ij})$ by a cutoff function,
$c_g(r_{ij})$.  A possible form for this cutoff function is
\[
c_g(r) = 
\left\{
\begin{array}{l@{\hspace{1em}}l}
1, & \mbox{for $r < r_{g1}$},\\
0, & \mbox{for $r \ge r_{g2}$},\\
\displaystyle
c_g(r_{g1}) \frac{r_{g2} - r}{r_{g2} - r_{g1}} & \mbox{otherwise},\\
\end{array}\right.
\]
with $r_{g1} \le r_{g2}$, which linearly tapers the energy to zero over
$[r_{g1},r_{g2})$.  Other tapering functions can be employed, or, by
choosing $r_{g2} = r_{g1}$, we can implement a sharp cutoff.  This type
of cutoff function implements a per-atom cutoff and is appropriate for
energy terms which are additive at large distances, such as the
Lennard-Jones potential.  The electrostatic potential, however, involves
substantial cancellation at large distances---two neutral molecules
interact via a dipole-dipole term which varies as $1/r^3$, while the
individual atom-atom terms decay as $1/r_{ij}$.  In this case, we need
to identify groups of atoms which should interact together.  The
residues of a protein provide a convenient grouping and we would
typically assign all the atoms in a small-molecule ligand to a single
group.  Compatible with the usage for a protein, we refer to these
groups as residues.  For each residue, $s$, we define a center position,
$\v b_s$, most conveniently defined as the center of mass, and a radius,
defined as the radius $h_s$ of the sphere centered at $\v b_s$ which
includes the van-der-Waals spheres of radius $\rho_i$ of all the
constituent atoms.  We then apply a ``per-residue'' cutoff function
multiplying the contribution from the residue pair $(s,t)$ by
$c_g(\abs{\v b_s - \v b_t} - (h_s + h_t))$.

The values used for the cutoff radii, $r_{g1}$ and $r_{g2}$, need to
evaluated based on the accuracy desired for the simulation.  This can be
determined by numerically determining the difference in the results
(either for the energies directly or for some derived quantity such as
binding affinity) between the finite- and infinite-cutoff energies.  In
applications to Monte Carlo codes, it is possible to carry out the
sampling at an energy approximating the actual energy and to compensate
for this when performing the canonical averages (which might be carried
out on a subset of the Markov chain).  In this case, the sampling energy
might entail using shorter cutoffs than would be warranted on the basis
of accuracy.  Having determined suitable cutoffs, it is a simple matter
to evaluate the energy avoiding treating atom pairs beyond the
respective cutoffs.  In the following, we treat electrostatic ($e$)
interactions, with a per-residue cutoff, and Lennard-Jones ($l$)
interactions, with a per-atom cutoff; furthermore we assume that
$r_{e2}\ge r_{l2}$, i.e., the electrostatic interactions are longer
range than the Lennard-Jones.

We first loop over all the atoms in each residue computing $\v b_s$ and
$h_s$ for all residues $s$.  We then loop over all pairs of residues, $s
\le t$, skipping any pair whose atoms all belong to the same fragment or
those for which $\abs{\v b_s - \v b_t} \ge r_{e2} + h_s + h_t$.  If the
residue pair survives these tests, then all atom pairs $(i,j)$ from
different fragments are considered; if $s=t$, we restrict the pairs to
$i<j$.  All such pairs contribute to the electrostatic energy while
those which satisfy $r_{ij} < r_{l2}$ contribute to the Lennard-Jones
energy.  There obviously is scope for additional optimization here.  For
example, the inner atom loop can be skipped if the second residue
belongs to a single fragment which matches the fragment of a particular
atom in the first residue.

Because of the way in which the cutoffs are applied, the result for the
energy is independent of the assignment of atoms to residues for energy
terms which use a per-atom cutoff.  In addition, {\em differences} in
the non-bonded energies are independent of the assignment of atoms to
fragments.  The energies for assemblies of 3 or more molecules can be
expressed in terms of the energies of 1 or 2 molecules.  These provide
useful checks on the implementation.

In some contexts it is useful also to define a ``steric'' energy term
which is infinite if any atoms overlap (with some definition of a
``hard'' atom radius) and is zero otherwise.  This provides a rapid
check of new configurations---particularly when trying to ``insert'' a
molecule during a grand canonical simulation \cite{adams75} or when
switching systems using the wormhole method \cite{karney05a}.  A
conservative definition of the hard atom radius is $0.55 \rho_i$ for
non-bonded atom pairs and $0.45 \rho_i$ for 1-4 atom pairs.  We skip the
check for 1-2 and 1-3 pairs and for those atoms with $\rho_i = 0$.  This
energy term can be implemented in essentially the same way as described
above but with scope for additional speedups.  The cutoff radius in the
residue-residue distance check can be replaced by 0.  An additional
atom-residue distance check can be be used to avoid executing the inner
atom loop if the outer atom is outside the sphere for the second
residue.  Finally, as soon as an overlap of hard spheres is detected the
routine can immediately return an infinite result.

\section{Implicit solvent models}

We now turn to the computation of the energy term for implicit solvent
models.  We focus here on the generalized Born solvent models
\cite{still90} and we have considered various implementations
\nocite{qiu97}\nocite{hawkins95}\nocite{hawkins96}%
\nocite{tsui00}\nocite{tsui01}\nocite{onufriev04}%
\cite{qiu97,hawkins95,hawkins96,tsui00,tsui01,onufriev04}.  Evaluating
the solvation energy for a system of molecules with such models is
typically orders of magnitude slower than computing the energy of the
molecules in vacuum.  The computation time is frequently compared to the
time to compute the energy with an explicit solvent model (including
$O(10^3)$ solvent molecules).  However, such comparisons are misleading
because implicit solvent models do not attempt to compute the energy of
a particular configuration of solvent molecules but to compute the {\em
free energy} of solvation, i.e., to average over all possible solvent
configurations for a given configurations of solute molecules.  Thus the
chief benefit of an implicit solvent model is to reduce dramatically the
number of degrees of freedom in the problem.  In the generalized Born
solvent models, the energy is written as the sum of two terms: a polar
term which is usually called the ``GB'' term and a cavity term which is
proportional to the solvent accessible surface area, the ``SA'' term.

The GB term involves long-range interactions and is the most costly to
compute.  We address the calculation of this term first.  The basic
expression is \cite{still90}
\begin{equation} \label{gpol}
G_\mathrm{pol} = -\frac12\frac1{4\pi\epsilon_0}
\biggl(1-\frac{\epsilon_0}{\epsilon_s}\biggr)
\sum_{i,j}q_iq_jf(r_{ij},\alpha_i,\alpha_j),
\end{equation}
where $\epsilon_s$ is the permittivity of the solvent,
$f(r_{ij},\alpha_i,\alpha_j)= [r_{ij}^2 +
\alpha_i\alpha_j\exp(-r_{ij}^2/(4\alpha_i\alpha_j))]^{-1/2}$, and the
double sum runs over {\em all} pairs of atoms (including $i=j$ and
$i\lessgtr j$).  In eq.~(\ref{gpol}), $\alpha_i$ is the ``generalized''
Born radius of the $i$th atom, which is larger that the ``bare'' Born
radius to account for the fact that atoms close to $i$ partially shield
it from the solvent.  $G_\mathrm{pol}$ represents the electrostatic
energy required to solvate a pre-assembled group of molecules and thus
this term is added to the vacuum electrostatic energy.  The various
implementations for the GB term differ in how $\alpha_i$ is computed.

For illustrative purposes, let us consider the model of Hawkins {\em et
al.}~\cite{hawkins95,hawkins96,tsui00,tsui01}.  (With minor
modifications, the technique is applicable to other GB models.)  We
express $\alpha_i$ as \cite[eq.~(10)]{hawkins95}
\begin{equation}\label{alpha}
\frac1{\alpha_i} = \frac1{\rho_i} -
 \sum_{j\ne i} \Delta_{ij},
\end{equation}
where $\rho_i$ is the radius of atom $i$,
\begin{equation}\label{deltaij}
\Delta_{ij}
  = \int_{\rho_i}^\infty \frac{dr}{r^2} H_{ij}(r; r_{ij}, \rho_j)
\end{equation}
is the reduction in the effective inverse Born radius of atom $i$ due to
atom $j$.  Here $H_{ij}$ is the fraction of the area of a sphere of
radius $r$ centered on the $i$th atom eclipsed by a $j$th atom and is
given by \cite[eq.~(12)]{hawkins95}
\[
H_{ij}=\left\{
\begin{array}{l@{\hspace{1em}}l}
\displaystyle
\frac{\rho_j^2-(r_{ij}-r)^2}{4r_{ij}r}, &
\mbox{for $\abs{r_{ij}-\rho_j} \le r \le r_{ij}+\rho_j$},\\
1, & \mbox{for $r < \rho_j - r_{ij}$},\\
0, & \mbox{otherwise ($r \gtrless r_{ij} \pm \rho_j$)}.\\
\end{array}
\right.
\]
Evaluating the integral in eq.~(\ref{deltaij}) then gives
\[
\Delta_{ij} = 
\left\{
\begin{array}{l@{\hspace{1em}}l}
0, & \mbox{for $ \rho_i > \rho_j + r_{ij}$},\\[1ex]
\displaystyle
\frac{l_{ij} - u_{ij}}2
-\frac{(r_{ij}^2-\rho_j^2)(l_{ij}^2-u_{ij}^2)}{8r_{ij}}\hspace{-10em}\\
\displaystyle
\hspace{3em}{}-\frac{\ln(l_{ij}/u_{ij})}{4r_{ij}} + l'_{ij},
 & \mbox{otherwise},
\end{array}
\right.
\]
where $u_{ij} = 1/(r_{ij}+\rho_j)$, $l_{ij} =
1/\max(\rho_i,\abs{r_{ij}-\rho_j})$, and $l'_{ij} =
1/\rho_i-1/\max(\rho_i,\rho_j-r_{ij})$.  The term $l'_{ij}$ is only
non-zero for $\rho_j > \rho_i + r_{ij}$, which is a possibility not
considered in \cite{hawkins95}.

Clearly $G_\mathrm{pol}$ is no longer the sum of pairwise atom-atom
contributions because the interaction of two atoms is affected by the
modification of the dielectric environment by a third atom.  However
$G_\mathrm{pol}$ may be evaluated by two pair-wise operations carried
out in sequence.  The first evaluates the generalized Born radii
$\alpha_i$ and the second computes the resulting electrostatic energy.

As with the treatment of the electrostatic and Lennard-Jones terms, we
can seek to limit the computational cost of evaluating $G_\mathrm{pol}$
by the use of cutoff functions.  Because eq.~(\ref{gpol}) provides the
dielectric screening for the vacuum electrostatic term, it is important
that the cutoff function multiplying $f(r_{ij},\alpha_i,\alpha_j)$
exactly match that used for the electrostatic term.

We also introduce a cutoff in eq.~(\ref{alpha}) by multiplying
$\Delta_{ij}$ by $c_b(r_{ij})$.  A per-atom cutoff is justified since
all the $\Delta_{ij}$ are positive.  Because $\Delta_{ij}$ scales as
$r_{ij}^{-4}$ for large $r_{ij}$, the error introduced by $c_b(r_{ij})$
scales relatively slowly as $r_{b1}^{-1}$.  In practice, this means we
need to make $r_{b1}$ reasonably large which in turn means that the cost
of evaluating $G_\mathrm{pol}$ in the case of a small ligand interacting
with a protein is much larger than the cost for the electrostatic
potential.  In particular, the screening of the ligand may modify the
Born radii of a large number of protein atoms and this unavoidably leads
to a large number of pair contributions to eq.~(\ref{gpol}).

The procedure for computing the energy outlined in the previous section
can now be modified to deal with the evaluation of $G_\mathrm{pol}$.  As
before our ``zero'' energy is given by separating all the fragments of
all the molecules infinitely far apart.  We set up the calculation of a
system of molecules by pre-computing $\alpha_{i0}$ which is given by
eq.~(\ref{alpha}) with the sum restricting to include only the
intra-fragment contributions (i.e., index $j$ ranges only over atoms
within the same fragment as atom $i$).  We compute $\Delta_{ij}$ and
$\Delta_{ji}$ together because they involve many of the same terms,
allowing the loops to be restricted to $i < j$, and we apply the Born
cutoff to the calculation of $\alpha_{i0}$.

When computing the energy of a molecular system, we compute all the
updates to the Born radii due to atoms in different fragments within the
Born cutoff, applying the same techniques of lumping the atoms into
residues described above (which allows the cutoff criteria to be applied
to groups of atoms) and of restricting the loops to $s \le t$ and, for
$s = t$, to $i < j$.  During this phase we mark all the residues which
contain atoms with $\alpha_i \ne \alpha_{i0}$.  We then make a second
pass over the atoms to evaluate the terms in eq.~(\ref{gpol}).  We use
the $i\rightleftharpoons j$ symmetry of the summand to make the
restrictions $s \le t$ and, for $s = t$, $i \le j$.  In the innermost
loop, we accumulate $q_iq_j f(r_{ij},\alpha_i,\alpha_j)$ if $i$ and $j$
belong to different fragments.  Otherwise, we add $q_iq_j
[f(r_{ij},\alpha_i,\alpha_j) - f(r_{ij},\alpha_{i0},\alpha_{j0})]$ and
we can skip this evaluation if both $\alpha_i = \alpha_{i0}$ and
$\alpha_j = \alpha_{j0}$.  In addition, we can skip pairs of residues if
all the atoms in each residue belong to the same fragment and if neither
residue is marked as having modified Born radii.

Salt effects \cite{srinivasan99} are easy to include within this
framework.  A minor complication occurs in the GB model of Qiu {\em et
al.}~\cite{qiu97} because $\alpha_{i0}$ depends on the ``volume'' of the
atoms and in this model the volume depends on the 1-2 bonded atoms which
may belong to a different fragment.  We account for this by assuming the
presence of such bonded atoms with an ideal bond length.  This is,
therefore, only exact if the inter-fragment bonds are at their ideal
lengths.  Our treatment here may be considered as a generalization of
the frozen atom approximation for GB/SA \cite{guvench02}.  However, in
our application we make all the approximations in the energy function
and the resulting energy is then a ``state variable'' and simulations
based on this are well behaved.  In contrast the implementation of
frozen atom approximation defines the energy so that it depends on the
history of the system which may cause the simulation to exhibit
unphysical properties.

\section{Solvent accessible surface area}

The other important contribution to the solvation free energy is the
cavity term.  This is obtained by placing spheres centered at each atom
with radius $a_i = \rho_i + r_w$ where $r_w$ is a nominal water radius
(typically $r_w = 0.14\,\mathrm{nm}$).  The cavity term is given by
\[
G_\mathrm{cav} = \sum_i \sigma_i A_i,
\]
where $A_i$ is the ``solvent accessible surface area'' for the $i$th
atom, i.e., the exposed surface area of the spheres around $i$ which is
not occluded by any other spheres and $\sigma_i$ is the surface tension
for the $i$th atom.  (Typically $\sigma_i$ is taken to be a constant
independent of atom, $\sigma_i \approx 3\,
\mathrm{kJ\,mol^{-1}\,nm^{-2}}$; however the method we describe does not
require this assumption.)  As before, the zero energy state is obtained
by separating the fragments infinitely.  The energy is then given by the
additional occlusion of the surface that occurs as the fragments are
assembled into molecules and the molecules brought into contact with one
another.

The exact evaluation of this term is quite complex and for this reason a
simple pairwise approximation has been developed \cite{weiser99}.
However, the errors in this method are poorly quantified.  This together
with the fact that this term is typically small compared to the
electrostatic terms in the energy lead us to develop a simple
zeroth-order quadrature method.  We select an accuracy level for the
cavity calculation $\delta$, e.g., $\delta = 0.1\,\mathrm{kJ/mol}$.  We
prepare for the calculation of the cavity term by placing each fragment
in a ``template'' position and we arrange a set of points on a sphere of
radius $a_i$ around each atom $i$.  The number of points is chosen to be
$N_i = \lceil 4\pi a_i^2 \sigma_i/\delta \rceil$.  The points are
distributed approximately uniformly around each sphere and the entire
surface energy of the sphere, $4\pi a_i^2 \sigma_i$ is divided among the
$N_i$ points.  (We will discuss the details of how to select the points
and assign the energy later.)  We next perform the intra-fragment
occlusion by deleting all the points of atom $i$ which are within $a_j$
of some atom $j\ne i$.  In this way each fragment is surrounded by a
cloud of surface points each representing about $\delta$ of cavity
energy.

In order to compute the cavity term for a particular molecular
configuration we transform the surface points for each fragment from
their template positions to their actual positions and make a copy of
the cavity energies for each point.  We consider all pairs of atoms
$(i,j)$ such that $i$ and $j$ are in different fragments and $r_{ij} <
a_i + a_j$.  We subtract from $G_\mathrm{cav}$ the energies of all the
points on atom $i$ that are within $a_j$ of atom $i$ and we set the
energies of these points to zero (to avoid their being counted multiple
times).  The optimizations described above can be used: the application
of a residue-residue cutoff (excluding residue pairs $(s,t)$ with
$\abs{\v b_s - \v b_t} \ge 2 r_w + h_s + h_t$), an atom-residue cutoff,
and the treatment of the $(i,j)$ and $(j,i)$ terms together.

In practice, the cost of evaluating this term is small for $\delta
\approx 0.1\,\mathrm{kJ/mol}$.  The error is proportional to $\delta$
and it is easy to benchmark a particular calculation by repeating it
with smaller $\delta$.  The resulting $G_\mathrm{cav}$ is obviously a
discontinuous function of configuration, jumping by $\pm \delta$ as
points move in and out of the water spheres of other atoms.  Thus it's
an inappropriate model for a molecular dynamics simulation.  However, it
yields satisfactory results for Monte Carlo simulations.

Let us return to the question of how to position the points on the atom
sphere and how to divide the energy between these points.  Ideally, we
would divide the energy of the sphere based on the area of Voronoi
polygons around each point.  The error will then be proportional to the
maximum radius of the Voronoi polygons and the ideal distribution of
points is the one which minimizes this maximum radius.  This is the
so-called ``covering problem'' for the sphere, i.e., how to cover a
sphere with identical discs \cite{fejestoth64}.  Unfortunately, there
are no general solutions to this problem.  So instead we divide the
sphere into equal intervals of latitude and we divide each latitudinal
interval longitudinally into approximately square regions.  A point is
placed at the center of each region and the area of the region is
assigned to that point.  Within each fragment, we alter the position of
the pole from one atom to the next, in order to avoid the occlusion of
many points simultaneously as fragments move relative to one another.

\section{Discussion}

We have shown how to make Monte Carlo moves for a molecular system with
constraints.  Constraints are imposed in a realistic way ensuring that
we obtain the right distribution corresponding to a thermodynamic
equilibrium.  We will still need to know this constrained distribution
if we wish to make wormhole moves \cite{karney05a}, because, in order to
satisfy detailed balance, we require knowledge of the wormhole volumes
and these include a factor proportional to the ``thickness'' of the
constraint manifold.  The adiabatic move involves, naturally, many
evaluations of the constraint energy raising a concern that the
implementation will be slow.  In reality, the cost of evaluating the
constraint energy is minuscule, particularly in comparison with the
solvation energy, so it is possible to evaluate the constraint energy
many thousands of times in the course of an adiabatic move with minimal
impact on the overall running time.  The method avoids much of the
algebra associated with other ways of imposing constraints
\cite{go70} and thus is more flexible and is easier to implement.

In the simple case of a molecule in which only a number of dihedral
angles are allowed to vary, the movement of all the atoms in the
molecule is bounded and thus the soft-energy acceptance probability is
reasonably large.  In contrast, the method where the dihedral angles are
perturbed may lead, due to a lever effect, to large motions if the
molecule itself is large.

This method can easily be generalized to do localized movements.  Thus,
we can tailor the random displacements of a protein to explore the
movement of a single loop.  Detailed balance is ensured if the random
displacement is a function of the atom but not of its position.  (The
general case can be accommodated by a suitable factor in the acceptance
probability.)  This method of localized movements is more widely
applicable than techniques such as ``concerted rotations''
\nocite{go70}\nocite{dodd93}\cite{go70,dodd93,mezei03}.  Artificially
fixing the positions of some atoms would, of course, mean that the moves
would not be ergodic.  This would be justified if we were interested in
examining the restricted system and we would then require ergodicity
over the restricted configuration space.

We have also considered how to optimize the evaluation of the energy in
a system of molecules made up of rigid fragments bonded together.  This
allows the use of implicit solvent at an acceptable cost.  If the system
is further constrained to allow only the variation of the torsion angle
of the inter-fragment bonds (fixing the bond lengths and bond angles),
then we should also consider modifying the force field to ``loosen'' the
torsion energies to counteract the effect of the hard constraints on the
other bond terms.  G\=o and Scheraga \cite{go69} show the importance of
considering such an effect and Katrich {\em et al.}~\cite{katritch03}
have offered a prescription for converting a general force field to
include this effect.  Alternatively, we might consider re-parameterizing
the torsion terms by carrying out constrained geometry optimizations of
model molecules where the energy of the molecule is minimized with the
dihedral angles fixed \cite{schmidt93}.

\section*{Acknowledgment}

This work was supported by the
\hrefx{https://mrmc-www.army.mil}{U.S. Army Medical Re-}{search and Materiel
Command} under Contract No.\ DAMD17-03-C-0082.  The views, opinions, and
findings contained in this report are those of the author and should not
be construed as an official Department of the Army position, policy, or
decision.  No animal testing was conducted and no recombinant DNA was
used.

\bibliography{free}
\end{document}